# Primordial Rotation of the Universe and Angular Momentum of a wide range of Celestial Objects

*C Sivaram*

Indian Institute of Astrophysics, Bangalore, 560 034, India

Telephone: +91-80-2553 0672; Fax: +91-80-2553 4043

e-mail: sivaram@iiap.res.in

*Kenath Arun*

Christ Junior College, Bangalore, 560 029, India

Telephone: +91-80-4012 9292; Fax: +91-80- 4012 9222

e-mail: kenath.arun@cjc.christcollege.edu

**Abstract:** The origin of rotation or spin of objects, from stars to galaxies, is still an unanswered question. Even though there are models which try to explain this, none of them can account for the initial impulse that gave rise to this spin. In this paper we present that a cosmological model that contains a term involving the primordial spin of the universe can explain how these objects acquired the property of spin. This model also gives a natural explanation for the quadratic scaling of angular momentum with mass. Again, from this model, the background torsion due to a universal spin density not only give rise to angular momenta for all structures but also provide a background 'centrifugal term' acting as a repulsive gravity accelerating the universe, with spin density acting as effective cosmological constant.

One of the biggest unanswered questions in astrophysics is "what is the origin of the rotation (angular momentum) or spin of all objects: from galaxies, stars, star clusters, etc.; and are they related to basic quantum properties of elementary particles?" Currently there are elaborate models that attempt to explain the angular momentum of objects ranging from planets to clusters of galaxies; however, none explain where the initial impulse moment comes from. The same dilemma applies for the spin of all objects – stars, galaxies, etc. (Trimble, 1984; Sivaram & Arun, 2011)

One possible explanation as to how all objects acquired the property of spin could be cosmological models which also contain a term involving the primordial spin of the universe. In homogenous and isotropic models, universe with matter may not only expand but also rotate (Gamow, 1946; Gödel, 1949; Barrow et al, 1985) (relative to local gyroscope). A general solution including the rotation is given by:

$$\frac{\dot{R}^2}{R^2} - \frac{8\pi G\rho}{3} + \omega^2 = 0 \qquad \ldots (1)$$

For the last two terms to be comparable:

$$\frac{8\pi G\rho}{3} = \omega^2 \qquad \ldots (2)$$

This implies a primordial angular frequency of: $\omega_o = 2 \times 10^{-18} Hz$ and a corresponding time period of:

$$T = 3 \times 10^{18} s \sim 10 T_H \qquad \ldots (3)$$

where $T_H = 10^{10}$ years is the Hubble time.

The angular momentum given by $J = M\omega R^2$, is conserved. And this cosmic rotation can possibly impart rotation to galaxies, clusters, stellar systems, etc. as we shall see below.

Recent work on the study of thousands of spiral galaxies imaged by Sloan Digital Sky Survey does in fact indicate that the universe could be spinning (Longo, 2011).

For a galaxy, the angular momentum is given by:

$$J_{gal} = M_{gal}\omega_o R_{gal}^2 \qquad \text{... (4)}$$

where, $M_{gal} = 10^{45} g$; $R_{gal}^2 = 10^{47} cm^2$ and we assume the same primordial value of angular frequency; $\omega_o = 2 \times 10^{-18} Hz$

This implies a galactic angular momentum of:

$$J_{gal} \approx 10^{74} ergs.s \sim 10^{100} \hbar \qquad \text{... (5)}$$

which is the observed angular momentum of galaxies!

Similarly for clusters of galaxies, the angular momentum is given by:

$$J_{Clust} = M_{Clust}\omega_o R_{Clust}^2 \approx 10^{110} \hbar \qquad \text{... (6)}$$

This is the observed angular momentum of large galaxy cluster.

(on a Hubble scale we have a corresponding $J \approx 10^{120} \hbar$)

In the case of the sun, the angular momentum is given by:

$$J_{Sun} = M_{Sun}\omega_o R_{Sun}^2 \approx 10^{50} ergs.s \qquad \text{... (7)}$$

which is only about 1% of the angular momentum of the solar system.

Considering the solar system we have:

$$J_{SolSys} = M_{SolSys}\omega_o R_{SolSys}^2 \approx 10^{52} ergs.s \qquad \text{... (8)}$$

which again matches observation.

The idea is that the initial solar nebula which condensed into the sun and planetary system also had its total angular momentum derived from the primordial rotation involving the same value of $\omega_o$.

Therefore we see that invoking primordial cosmic rotation can give rise to the observed rotation angular momenta of galaxies, galaxy clusters, stellar planetary systems, etc., the origin of which is otherwise not clearly understood (Gamow, 1946; Barrow et al, 1985). It is remarkable that the same (universal) primordial value of $\omega_o$ was used in all the above cases.

Including the $\Lambda$ term the general solution is given by:

$$\frac{\dot{R}^2}{R^2} - \frac{8\pi G\rho}{3} + \omega^2 + \frac{\Lambda c^2}{3} = 0 \qquad \text{... (9)}$$

The rotation can produce an acceleration analogous to that of DE. This acceleration is given by:

$$\omega^2 R_{Univ} \approx 10^{-7} cms^{-2} \qquad \text{... (10)}$$

which is exactly the expected value of acceleration required to explain the accelerated expansion:

$$\ddot{R} = \frac{4\pi G\rho R}{3} \sim 10^{-7} cms^{-2} \qquad \text{... (11)}$$

For the DE term to be comparable to the rotation we have:

$$\frac{\Lambda c^2}{3} = \omega^2 \Rightarrow \Lambda = \frac{3\omega^2}{c^2} \sim 10^{-56} cm^{-2} \qquad \text{... (12)}$$

which is precisely the present observed value. Thus a cosmological model with a large scale primordial rotation term of this order can give an accelerating universe mimicking a dark energy term.

Even on a cosmic scale of $\approx 10^{25} cm$, the deviation from isotropy will be of the order of $\sim \left(\frac{10^{25}}{10^{28}}\right)^2 \sim 10^{-6}$, which is one order less than the sensitivities of the current probes $10^{-5}$. Even such a primordial rotation if it exists would be barely detectable.

Since the angular momentum $J$, is conserved, rotation will be dominant during the radiation era, since both $\omega^2$ and $\rho_{rad}$ will fall off as $1/R^4$, where as in matter dominated era, $\omega^2$ falls off at a faster rate.

For a whole hierarchy of objects to be gravitationally bound, their gravitational (binding) self energy density should be at least equal or exceed the background cosmological gravitational self energy density (which again equals the critical matter density) (Sivaram, 1993; 1994; 2000; 2005; 2008). This implies that, from the equation:

$$\frac{GM^2}{R^4} \approx \frac{\Lambda c^4}{8\pi G} \quad \text{... (13)}$$

$$\frac{M}{R^2} = \frac{c^2}{G}\sqrt{\Lambda} \quad \text{... (14)}$$

That is:

$$R^2 = \frac{GM}{c^2}\frac{1}{\sqrt{\Lambda}} \quad \text{... (15)}$$

The angular momentum is then given by:

$$J = M\omega R^2 = M\omega_o \frac{GM}{c^2}\frac{1}{\sqrt{\Lambda}} \quad \text{... (16)}$$

Thus (Wesson, 1983):

$$J \propto M^2 \quad \text{... (17)}$$

This relation, as pointed out earlier in several papers (Sivaram, 1987; de Sabbata & Sivaram, 1988) holds for a very wide range of celestial objects. This relation has often been difficult to understand although it is empirically well established. (Sivaram, 1984)

Here apart from getting reasonable values for the angular momenta of a wide range of objects we have a natural explanation for this quadratic scaling with mass. The value of the constant in the proportionality relation is related to $\omega_o$ above and works out to the right order of magnitude (~$10^{-17}$).

Also the spin densities ($\sigma$ = spin/volume) for a range of objects work out to be the same. For an electron, the classical radius is $\sim 3\times 10^{-13} cm$, and the spin density is given by:

$$\sigma_e = \frac{0.5\hbar}{\left(\frac{4}{3}\pi r_e^3\right)} \sim 10^9 \, ergs.s/cc \quad \text{... (18)}$$

Similarly for a proton: $\sigma_P \sim 10^9 \, ergs.s/cc$ ... (19)

Also for the solar system we have: $\sigma_{SolSys} \sim 10^9 \, ergs.s/cc$ ... (20)

For a galaxy, the angular momentum is (from equation (5)) $J_{gal} \sim 10^{100}\hbar$, the spin density is then given as:

$$\sigma_{gal} = \frac{10^{100}\hbar}{\left(\frac{4}{3}\pi R_{gal}^3\right)} \sim 10^9 \, ergs.s/cc \qquad \ldots (21)$$

And for the universe, $J_{Univ} \sim 10^{120}\hbar$, and the corresponding spin density is then given as:

$$\sigma_{Univ} = \frac{10^{120}\hbar}{\left(\frac{4}{3}\pi R_H^3\right)} \sim 10^9 \, ergs.s/cc \qquad \ldots (22)$$

From equations (18) – (22), we see that within an order of magnitude spin density i.e. spin or angular momentum per unit volume is same for all structures, from elementary particles to the universe. The spin density in Einstein-Cartan theory is related to torsion ($Q$) of background space-time as:

$$Q \propto \sqrt{\text{curvature}} \qquad \ldots (23)$$

The torsion is given by:

$$Q = \frac{4\pi G \sigma}{c^3} \qquad \ldots (24)$$

where the spin density:

$$\sigma = \frac{J}{R^3} \qquad \ldots (25)$$

and $R$ is the scale size of background space.

From equations (16) – (20), we have: $\sigma = 10^9 \, ergs.s/cc$ $\qquad \ldots (26)$

Therefore the torsion is given by:

$$Q = \frac{4\pi G \sigma}{c^3} \sim 10^{-28} \, cm^{-1} \qquad \ldots (27)$$

And the background curvature $\sim Q^2 \approx 10^{-56} \, cm^{-2}$ $\qquad \ldots (28)$

Torsion gives a modification of Poisson equation as: (de Sabbata & Sivaram, 1994)

$$\nabla^2 \phi = 4\pi G(\rho - G\sigma^2) \qquad \ldots (29)$$

and acts opposite to gravity and is a repulsive term. Indeed, the $G\sigma^2$ term (for a constant $\sigma$ as implied by equations (18-22)) gives rise to a potential $\sim G\sigma^2 r^2$, growing with $r$ like a cosmological constant term. This is precisely the order of the observed $\Lambda$.

Hence it could be a possible candidate for dark energy!

At present epoch, $\dfrac{G\sigma^2}{c^4}(\sim 10^{-29} g/cc) > \rho$, where $\rho \sim 3 \times 10^{-30} g/cc$.

The spin density in Cartan's equation plays the role of a repulsive potential. The non-minimal spin-spin interaction term arising out of the interaction of the interaction of a Dirac spin or particle with the vierbein gravitational field gives rise to an "effective cosmological constant term", as indicated in (Isham, Salam & Strathdee, 1973; Sivaram, Sinha & Lord, 1974; 1975).

Again it has been shown that the so called teleparallel gravity (first introduced by Einstein as a possible alternative to GR) which uses the torsion instead of the curvature in the gravitational action is equivalent to a theory with a cosmological constant. (de Sabbata & Sivaram, 1994)

Recent studies (Kashlinsky, et al, 2008) on the fluctuations in the cosmic microwave background have revealed a coherent bulk flow of clusters of galaxies on a large cosmic scale. This flow (termed the dark flow) is difficult to explain with the current cosmological models. But invoking a primordial rotation with the same value of $\omega_o$ can account for the observed peculiar velocities (which on the scales of $\sim 300 Mpc$ could give rise to coherent bulk flows of $\sim 3 \times 10^4 km/s$, suggesting increase of large scale velocities with distance) of these clusters of galaxies.

So the background torsion due to a universal spin density $\sigma$ can not only give rise to angular momenta for all structures but also provide a background 'centrifugal term' acting as a repulsive gravity accelerating the universe, mimicking the effective cosmological constant or accelerating dark energy of the observed magnitude.